\documentclass[format=sigconf, screen=true, review=false]{acmart}
\setcopyright{rightsretained}
\acmPrice{}
\acmDOI{10.1145/3611643.3613095}
\acmYear{2023}
\copyrightyear{2023}
\acmSubmissionID{fse23demo-p26-p}
\acmISBN{979-8-4007-0327-0/23/12}
\acmConference[ESEC/FSE '23]{Proceedings of the 31st ACM Joint European Software Engineering Conference and Symposium on the Foundations of Software Engineering}{December 3--9, 2023}{San Francisco, CA, USA}
\acmBooktitle{Proceedings of the 31st ACM Joint European Software Engineering Conference and Symposium on the Foundations of Software Engineering (ESEC/FSE '23), December 3--9, 2023, San Francisco, CA, USA}
\received{2023-05-11}
\received[accepted]{2023-07-20}

\usepackage[T1]{fontenc}
\usepackage{amsmath,amsfonts}
\usepackage{microtype}
\usepackage{xspace}
\usepackage{listings}
\usepackage{textgreek}

\definecolor{dkgreen}{rgb}{0,0.6,0}
\definecolor{gray}{rgb}{0.5,0.5,0.5}
\definecolor{mauve}{rgb}{0.58,0,0.82}
\definecolor{shadecolor}{rgb}{0.95,0.95,0.95}

\definecolor{pblue}{rgb}{0.13,0.13,1}
\definecolor{pgreen}{rgb}{0,0.5,0}
\definecolor{pred}{rgb}{0.9,0,0}
\definecolor{pgrey}{rgb}{0.46,0.45,0.48}
\PassOptionsToPackage{usenames,dvipsnames}{color}
\PassOptionsToPackage{dvipsnames}{xcolor}

\lstdefinestyle{toplisting}{
  float=tp,
  floatplacement=tbp,
}

\AtBeginDocument{%
  \providecommand\BibTeX{{%
    \normalfont B\kern-0.5em{\scshape i\kern-0.25em b}\kern-0.8em\TeX}}}

\begin{document}
\title{Helion: Enabling Natural Testing of Smart Homes}

\author{Prianka Mandal}
\affiliation{
  \institution{William \& Mary}
  \streetaddress{}
  \city{Williamsburg}
  \state{Virginia}
  \country{USA}
  \postcode{}
}
\email{pmandal@wm.edu}
\orcid{0009-0009-1329-9818}

\author{Sunil Manandhar}
\affiliation{
  \institution{IBM Research}
  \streetaddress{Thomas J. Watson Research Center}
  \city{Yorktown Heights}
  \state{NY}
  \country{USA}
}
\email{sunil@ibm.com}
\orcid{0000-0003-3187-0044}

\author{Kaushal Kafle}
\affiliation{
  \institution{William \& Mary}
  \streetaddress{}
  \city{Williamsburg}
  \state{Virginia}
  \country{USA}
  \postcode{}
}
\email{kkafle@wm.edu}
\orcid{0000-0003-1917-7677}

\author{Kevin Moran}
\affiliation{%
  \institution{University of Central Florida}
  \city{ Orlando}
  \state{Florida}
  \country{USA}}
\email{kpmoran@ucf.edu}
\orcid{0000-0001-9683-5616}

\author{Denys Poshyvanyk}
\affiliation{
  \institution{William \& Mary}
  \streetaddress{}
  \city{Williamsburg}
  \state{Virginia}
  \country{USA}
  \postcode{}
}
\email{denys@cs.wm.edu}
\orcid{0000-0002-5626-7586}

\author{Adwait Nadkarni}
\affiliation{
  \institution{William \& Mary}
  \streetaddress{}
  \city{Williamsburg}
  \state{Virginia}
  \country{USA}
  \postcode{}
}
\email{apnadkarni@wm.edu}
\orcid{0000-0001-6866-4565}

\renewcommand{\shortauthors}{Mandal et al.}

  

\newcommand\codel[1]{\begin{verbatim}{#1}\end{verbatim}}

\newcommand\inline[1]{{\lstinline[keywordstyle=\color{black},basicstyle=\scriptsize\ttfamily,stringstyle=\color{black}]{#1}}}
\newcommand\inlinesmall[1]{{\lstinline[keywordstyle=\color{black},basicstyle=\small\ttfamily,stringstyle=\color{black}]{#1}}}

\newcommand\myparagraph[1]{\noindent{\bf {#1}:}}
\newcommand\myparagraphnew[1]{\noindent{\bf {#1}:}}

\newcommand{\arrow}{{$\rightarrow$}\xspace}
\newcommand{\ie}{\textit{i.e.,}\xspace}
\newcommand{\eg}{\textit{e.g.,}\xspace}
\newcommand{\etc}{\textit{etc.}\xspace}
\newcommand{\etal}{\textit{et al.}\xspace}
\newcommand{\etals}{\textit{et al.'s}\xspace}
\newcommand{\aka}{\texttt{a.k.a.}\xspace}
\newcommand{\mascengine}{\texttt{\tool{} Engine}\xspace}
\newcommand{\masclab}{\texttt{\tool{} Lab}\xspace}
\newcommand{\pluginmanager}{\texttt{Plugin Manager}\xspace}
\newcommand{\configurationmanager}{\texttt{Configuration Manager}\xspace}
\newcommand{\TODO}[1]{{\color{red}{\textbf{TODO: {#1}}}}}
\newcommand{\PRIANKA}[1]{{\color{blue}{\textbf{PRIANKA: {#1}}}}}
\newcommand{\REFH}[0]{{\color{red}REF HERE}}

\makeatletter
\newcommand{\linebreakauthor}{%
  \end{@IEEEauthorhalign}
  \hfill\mbox{}\par
  \mbox{}\hfill\begin{@IEEEauthorhalign}
}

\newcommand{\tool}{{H\textepsilon{}lion}\xspace}
\newcommand{\tools}{{H\textepsilon{}lion}'s\xspace}
\newcommand{\toolHA}{{H\textepsilon{}lionHA}\xspace}
\newcommand{\toolHAs}{{H\textepsilon{}lionHA}'s\xspace}
\newcommand{\ngram}{{n-gram}\xspace}
\newcommand{\sequence}[1]{{\sf \small #1}}
\newcommand{\routine}[2]{{\bf IF} {\sf \small #1} {\bf THEN} {\sf \small #2}}
\newcommand{\add}[1]{{{\xspace#1}}}

\newcommand{\homeassistant}{{\sf Home Assistant}\xspace}
\newcommand{\homeassistants}{{\sf Home Assistant}'s\xspace}
\newcommand{\REF}{{\color{red} \textbf{[REFS]}}\xspace}

\begin{abstract}
Prior work has developed numerous systems that test the security and
safety of smart homes. 
For these systems to be applicable in practice, it is necessary to test
them with realistic scenarios that represent the use of the smart home,
\ie home automation, in the wild.
This demo paper presents the technical details and usage of \tool, a system that uses n-gram language modeling
to learn the regularities in
user-driven programs, \ie routines developed for the smart home, and
predicts {\em natural} scenarios of home automation, \ie event sequences
that reflect realistic home automation usage.
We demonstrate the \toolHA platform, developed by integrating \tool with the popular \homeassistant platform.
\toolHA allows an end-to-end exploration of \tools scenarios by
executing them as test cases with real and
virtual smart home devices.

\noindent The demo video can be found here: \url{https://youtu.be/o9g0wKiJJMI}

\end{abstract}


\maketitle

\section{Introduction}
\label{sec:intro}

In smart home platforms, automation is driven by trigger-action programs known as {\em routines}, wherein a certain {\em action} event is programmed to occur after a certain {\em trigger}, \eg IF the user is home (trigger), THEN turn the camera off (action).
Prior work has analyzed routines created by developers (\ie  IoT apps) to understand the security and safety issues in home automation~\cite{cmt18,ctm19,jcw+17,whbg18,cbs+18, kmm+19, kmm+20}.
However, IoT apps defined by developers may not reflect realistic home automation use in the wild, \ie the events that are likely to actually occur in end-user homes.
The unavailability of realistic home automation usage makes it difficult to design or evaluate systems designed to analyze/test home automation in a practical manner.

For instance, consider the problem of testing the effectiveness of a security analysis/system for evaluating home automation.
At present, researchers generally evaluate their systems with random permutations of smart home events~\cite{jcw+17,ctm19, whbg18}, which may not represent realistic home automation usage in the wild, and lead to an impractical design or evaluation of the systems.
One possible solution to this problem would be to collect real execution traces of smart home events from end-user homes, and then use those traces to build/evaluate systems.
However, not only is this approach extremely privacy-invasive (\ie as the traces also represent physical events in the user's home), but may also be ineffective, since the traces may contain superfluous events that represent platform and device-specific intricacies, \ie "noise", which may distract from the real, semantically-relevant, smart home usage.
Thus, there is a need for synthetically generated but realistic home automation scenarios that can be used to generate effective test cases.

We previously built a framework, \tool~\cite{manandhar2020towards}, that leverages {\em user-driven routines}, \ie routines created by end-users using simple trigger-action user interfaces provided by most popular smart home platforms (\eg NEST~\cite{nest}, SmartThings~\cite{smartthings}). 
User-driven routines represent the real home automation requirements of users, as they allow end-users to express their home automation workflows/programs via the UI, without writing a single line of code, and hence, eliminating the need for (or relevance of) developer-provided IoT apps.
That is, routines obtained from a user, combined with simple cues regarding their order/frequency of execution, form the "home automation program" for that user.
\tool builds upon prior work in the SE domain on leveraging the naturalness in code for tasks such as code completion~\cite{Hindle:ICSE12}, and similarly, learns the regularities in a corpus of such home automation programs derived from end-users using \ngram LMs.
\tools model can then be used as a sequence generator to predict natural scenarios, \ie realistic home automation event sequences based on a given history of events.
These natural scenarios can then be used as test cases in the design and evaluation of security systems built for the smart home, \eg in lieu of random events used by prior work.
The full details of \tools methodology, design considerations, evaluation results, and discussion of the findings
are described in our past study~\cite{manandhar2020towards}. The source code of \tool is also publicly available on GitHub~\cite{onlineapp}.

This demo paper describes the implementation and usage of \tool, and particularly, the implementation of \toolHA, an extension to the popular \homeassistant platform~\cite{home+assistant} with \tool, which enables users to generate natural scenarios as well as automatically execute them as test cases with real/virtual devices connected to the platform.  
\toolHA first provides a UI for configuring \tools model and automatically generating scenarios adhering to the configuration.
These scenarios are in the form of sequences of event tokens predicted by \tool. 
\toolHA converts the event tokens from the sequences into \homeassistant-specific events, and passes the events on to the system by interfacing with relevant components (specifically, the Event Bus), and in this manner, {\em executes} the scenario on the platform.
\toolHA can be connected to physical devices, or virtual devices configured via the UI, to execute a wide variety of \tools scenarios. 
\toolHAs dashboard also enables the user to monitor various smart home states during the execution of a scenario.
The source code and documentation of \tool on \homeassistant, \ie\ \toolHA, are publicly available ~\cite{helion-on-home-assistant}.

The paper is organized as follows: 
Section~\ref{sec:background} discusses the necessary background on language modeling and the fundamentals of \homeassistant. 
Section~\ref{sec:helion} provides a brief overview of \tool.
Section~\ref{sec:implementation} presents the detailed design and implementation of \tool and Section~\ref{sec:helionha} presents the integration of \tool into \homeassistant, \ie\ \toolHA.
Finally, Section~\ref{sec:conclusion} concludes the paper.

\section{Background}\label{sec:background}
This section provides the rationale behind choosing the \ngram language model for \tool and a brief overview of Home Assistant. 

\subsection{N-gram Language Model}
Hindle \etal demonstrated that source code written by humans is just like natural language and thus, contains patterns that make it predictable~\cite{Hindle:ICSE12}. 
Similarly, user-driven routines are also natural as these are created by humans to fulfill a particular workflow, and hence, may contain regularities/patterns that are predictable. %
Based on this, we use statistical language modeling to learn the regularities in home automation sequences composed of user-driven routines (\ie which represent the end-user's overall ``program''), and use the model to predict natural scenarios of home automation. 

In general, language models (LMs) measure the probability of a sentence $s=w_1^m=w_{1}w_{2}...w_{m}$, given the probabilities of the individual words in the sentence (\ie $w_1^m$), as previously estimated from a training corpus.
This ability enables prediction, \ie predicting the next most probable word that can follow a sequence of words, known as the ``context'' or ``history''. 
When modeling smart home routines, we define a ``sentence'' to represent a sequence of home automation events, wherein the ``words'' (a.k.a {\em tokens}) are smart home events (\eg \sequence{$<$LightBulb, switch, ON$>$}).

In \tool, we specifically use \ngram LMs as they assume the Markov property, \ie instead of computing the conditional probability given an entire event or language history, we can approximate it by considering only a few tokens from the past as the history. 
The intuition behind n-gram LMs applied to natural language is that shorter sequences of words are more likely to co-occur in training corpora, thus providing the model with more examples to condition token probabilities, enhancing its predictive power.  
Using the \ngram model, we estimate the probability of the event sequence $s=e_1^m=e_{1}e_{2}...e_{m}$ as follows:
\begin{equation}
\label{eq:nLanguageModel}
p(e_1^m) = \prod_{i=1}^m p(e_i|e_1^{i-1}) \approx \prod_{i=1}^m p(e_i|e_{i-n+1}^{i-1})
\end{equation}

\subsection{Home Assistant}\label{subsec:ha}
Home Assistant is an open-source software platform for home control and automation~\cite{home+assistant}. 
Here we briefly describe the key areas of Home Assistant that are necessary for integrating \tool into it to develop \toolHA. 

\myparagraph{The \homeassistant dashboard} The dashboard is a customizable page where users can manage their home using HomeAssistant's mobile and Web interfaces. 
The overview dashboard is the first thing that users see after installing \homeassistant. 
The dashboard displays information connected to and available in Home Assistant, including the connected devices, both real and virtual.

\myparagraph{Cards} The \homeassistant Dashboard is composed of cards. 
Each card has its own configuration options, which users can configure as required. 
Moreover, users can build and use their own cards. 
One of the most common cards is the ``entities'' card which groups abstract items together into lists.

\myparagraph{Configuration} Other than the user interface, users can configure their Home Assistant instance by editing {\sf configuration.yaml}. The {\sf configuration.yaml} file contains integrations to be loaded along with their configurations.

\section{Helion}
\label{sec:helion}
\begin{figure}[t]
    \centering
    \includegraphics[width=.47\textwidth]{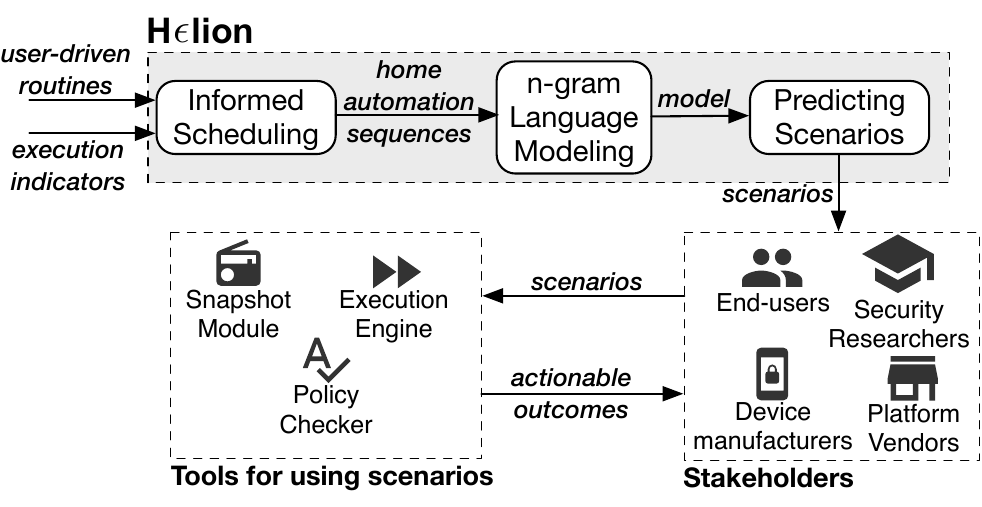} 
    \caption{{\small An overview of the \tool framework, which models home automation sequences to construct natural scenarios. Stakeholders use tools that analyze or execute scenarios to obtain actionable outcomes.}}
\label{fig:overview}
\end{figure}
Figure~\ref{fig:overview} shows \tool, a data-driven framework that models the regularities of user-driven home automation, generates natural home automation scenarios, and provides stakeholders with tools to use the scenarios and obtain actionable outcomes. 

The first step of \tool is \textbf{data collection}, \ie collecting {\em user-driven routines} from users along with {\em execution indicators}. 
Execution indicators are clues about when or how frequently those user-driven routines are scheduled to execute. 
The next step is \textbf{informed scheduling} where the routines and execution indicators are transformed into home automation {\em event sequences}, which represent the user's ``program'' for a certain duration (\eg a month). 
During the \textbf{modeling} phase, \tool uses n-gram LMs on the corpus of event sequences obtained from users.
Finally, during \textbf{scenario prediction}, \tool generates natural scenarios, which can be used by stakeholders for testing the smart home, such as executable {\em test cases} in \toolHA to be executed with real and virtual devices.

\section{Design and Implementation}
\label{sec:implementation}
This section describes the design and implementation of \tool, elaborating on its four steps: data collection and representation, informed scheduling, modeling, and scenario prediction. 
\subsection{Data Collection and Representation}
The most practical way to collect user-driven routines is collecting routines directly from users. To do this, we conducted a user survey with 40 users and obtained 273 routines (233 unique) created by the users. After collecting data from users, we transformed those into home automation tokens. Anonymized datasets and code for the \tool are available at~\cite{onlineapp}.

\myparagraph{Collecting Data from Users}
We used the survey to collect the routines from users. 
In the survey, participants were asked to create routines as in the ``IF'' and ``THEN'' trigger-action format, but expressed in plain English text, allowing users to express any functionality desired without enforcing artificial constraints. 
Here is the raw routine from our dataset:
\add{
\begin{center}
    \routine{the motion is detected}{camera takes a picture}
\end{center}
}

After creating routines, participants specified execution indicators, \ie the time-range, day-range, and frequency indicators for their routines (described in detail in Section~\ref{sec:informed-scheduling}). 

\myparagraph{Representing smart home events as tokens} A token is a home automation event parsed from a structured natural language routine. Here, an event can denote a change in the state of a device (\eg door locked) or the home (\eg the user is away). We define \tools home automation event token as:

\begin{center}
    $e_i :=<device_i, attribute_i, action_i>$
\end{center}

\noindent where $device_i$ represents the device (\eg door lock, camera), the $attribute_i$ corresponds to one of a predefined set of device attributes (\eg the {\em lock} attribute for the door lock, which can take the values {\sc Locked}/{\sc Unlocked}), and $action_i$ represents the change of state, and hence, the current value of $attribute_i$. 
\add{\noindent Using this design, the example routine discussed previously (\ie the motion sensor/camera) would be tokenized as (terms from SmartThings capabilities~\cite{smartthingsCaps}):
\begin{center}
\noindent {\centering $<motion\_sensor,\ motion,\ ${\sc detected}$>$, $<security\_camera,\  image,\ ${\sc take}$>$}
\end{center}
}

\subsection{Informed Scheduling}
\label{sec:informed-scheduling}
A home automation event sequence is an approximate ordered representation of how the routines would execute in the user's home in a particular timeframe. 
\tool transforms the tokenized routines specified by a particular user into a home automation event sequence.  Here, the order is important for generating home automation sequences. 
Therefore, we introduce a novel abstraction for users to stipulate the approximate order in which routines may execute, \ie routine-specific execution indicators. 
Users may be able to describe when they perform certain personal tasks which trigger home automation,\ie when they come home, go to work, bed, cook, or do laundry. 
Execution indicators allow us to capture such factors, which we then leverage to schedule routines to create home automation sequences. 
This is why we define the approach as informed scheduling, as the scheduling mechanism is informed by the user's understanding of their own home use.

Execution indicators constitute the time and frequency of the potential execution of a routine. 
For \tool, we consider three types of indicator: (1) the time-range indicator (\eg early morning, noon, and night), (2) the day-range indicator (\eg mostly on weekdays, and mostly on weekends), and (3) the frequency indicator (\eg many times a day, few times a day, few times a month). 
Routines collected from users have been scheduled in the time series based on these indicators, also provided by the users, we extract the ordered set of routines from this month-long series as the execution sequence and construct the {\sf HOME} corpus. 
The {\sf HOME} corpus consists of 30,518 events, from 40 month-long sequences (\ie 40 users), generated from 273 routines (233 unique) and their execution indicators.

\vspace{-.5em}
\subsection{Modeling}
\tool uses the \ngram LM to learn the regularities in user-driven home automation sequences. For the \tools n-gram model, we choose $n\ge3$. 
Choosing values of $n<3$ can either completely ignore the context or capture simple relationships that are already observable from data. 
Considering larger values of $n$, \ie $n\ge 3$, the model can learn non-obvious regularities in home automation corpora. 
However, considering too much of the event history (\ie a very large $n$) may actually hurt the predictive power of the model. Moreover, longer sequences may be relatively uncommon in the wild, even if they are realistic and useful for uncovering serious security/safety flaws. Therefore, choosing $n\ge3$ leads to a better model. 
We used interpolated smoothing methods since it performs well with lower-order (\ie 3-4 gram) models~\cite{chen-goodman}.
\vspace{-.8em}
\subsection{Scenario Prediction}
\tool considers the language model as a sequence generator that can produce an arbitrarily long series of events, \ie home automation scenarios. 

For security and safety-related testing, researchers require both natural scenarios that are reasonably likely to occur in end-user homes and unnatural scenarios that demonstrate unsafe situations. Hence, we designed \tool to generate two corresponding flavors of scenarios: 

\myparagraph{The {\em up} flavor, natural scenarios} This is the default flavor where \tool predicts the most probably event(s) given a history, \ie generates natural home automation scenarios.

\myparagraph{The {\em down} flavor, unrealistic scenarios} The down flavor generates unrealistic/unnatural scenarios by predicting the most improbable event(s) given a particular history.

\section{Implementation of \toolHA}\label{sec:helionha}
We implemented \toolHA by integrating \tool with \homeassistant. 
In this section, we describe how we implement \tool on the \homeassistant platform to develop \toolHA as well as how users can use it. 

\subsection{Conversion of \tool Tokens to Entities and Cards in \homeassistant}
Entities in \homeassistant are abstract objects that hold the state of the simulated device, \ie each entity is \homeassistants representation of the function of a device. 
To elaborate, \homeassistant allows users to connect to physical devices, and usually, entities only serve as the interface to those devices. 
As we defined in Section~\ref{sec:implementation}, a token is a \tools representation of a device, its attribute, and its state (\ie token, $e_i$ = $<device_i$, $attribute_i$, $action_i>$). 
To implement \toolHA, we needed to convert each token to an entity name. 
We implemented the \inlinesmall{parse\_token.py} script which takes in a token as an input and returns the corresponding entity. 
If the token contains multiple devices, then each entity name will be separated by a space. This entity is then added to the \inlinesmall{ui-lovelace.yaml} or \inlinesmall{helion.yaml} file.

We used two types of entities in \toolHA: (i) \inlinesmall{input\_boolean} for keeping track of two states (\eg lightbulb has two states: on or off). (ii) \inlinesmall{input\_select} for keeping track of multiple or complex states (\eg the motion sensor has four states: activated, deactivated, detected, and not\_detected).

Cards are the components that are displayed in the \homeassistant dashboard and represent entities. The state of the entity can be seen or changed through cards. 
In our implementation, each card is set up to correspond to an entity. 
Users can also input a token through an input card which will run a script to modify \inlinesmall{helion.yaml}. We implemented a script \inlinesmall{change\_ui\_cards.py}, that is used to create the cards that are displayed on the \homeassistant dashboard. When tokens are output from the \tool server, we take the tokens that are output and pass them to this script, which finds the corresponding card and modifies \inlinesmall{helion.yaml}. 

\begin{figure}[tbp]
    \centering
    \includegraphics[width=.47\textwidth]{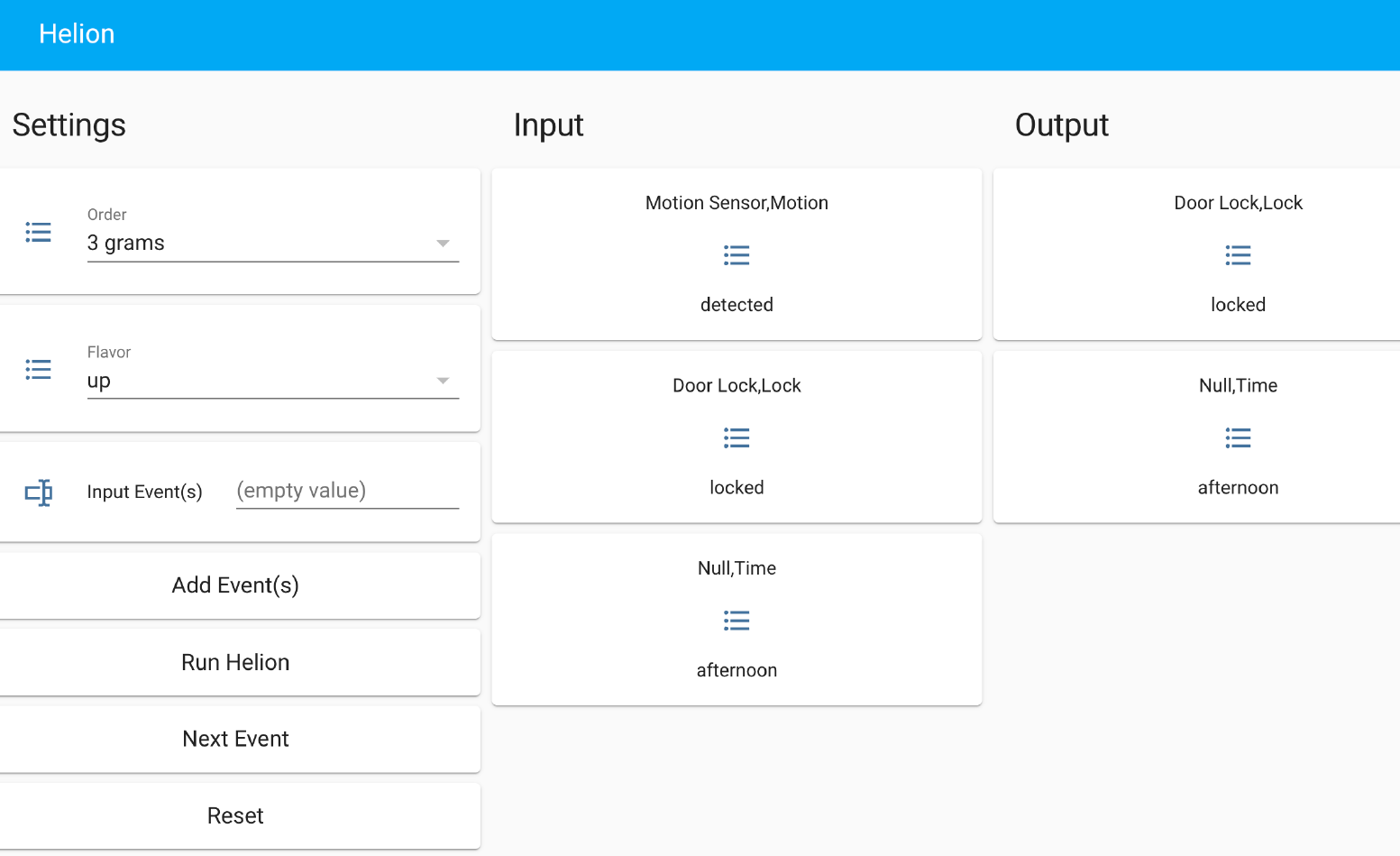} 
    \caption{{\small {An overview of the \tool on \homeassistant platform}}}\label{fig:demo}
    \vspace{-1em}
\end{figure}

\subsection{Predicting Scenarios using the \toolHA UI and Executing them as Test Cases}
In this section, we describe how exactly \tool integrates with \homeassistant, in terms of both the user interface (\ie the \toolHA dashboard) as well as the backend (\ie the \tool server). 

We created a custom page on \homeassistant where the UI specific to \tool resides. 
The new page is built on top of Lovelace UI which is a customizable \homeassistant dashboard~\cite{lovelace}. Figure~\ref{fig:demo} shows the UI of \toolHA. 
The UI of \toolHA has three main parts: Settings, Input, and Output, each consisting of several cards.  

The user first specifies the settings, particularly the order (3-gram or 4-gram), and flavor (up or down). 
These elements are stored as \inlinesmall{input\_select} entities in the backend. %
The user then specifies the event history, \ie the input events that form the context following which \tool predicts future events.
When ``Run Helion'' is clicked on the dashboard UI, \toolHA takes the user-provided history/input events from the \inlinesmall{input\_list} and reprocesses them to send to the \tool server (which executes in the backend).
The \tool server then provides predictions given the settings and inputs, and \toolHA transforms them into \homeassistant events, and displays the same in the dashboard.

To elaborate, the scenario, \ie predictions generated by the \tool server are gathered in up.tsv and down.tsv based on which flavor the user requested.
Note that these predicted events are \tools tokens, \ie in the form $<device, attribute, state>$, which must now be converted to \toolHA's events, to be executed with real and virtual devices on the platform.

To generate each \homeassistant event (represented as [device.attribute, state]) corresponding to a predicted token, we invoke \toolHAs$ $\inlinesmall{ call\_service} method, which takes the entity\_id (\ie the device) and the target\_state (\ie the state) as parameters, along with the method to call (\ie the attribute) to properly set the entity\_id to its target\_state. 
This is equivalent to sending a command to a physical (or virtual) device.  
In essence, after the call\_service method is invoked, the device.attribute in the \homeassistant token is set to the state specified within that token.
When the token is executed, it is broadcast as an event in \homeassistants Event Bus to let other entities (and automation) know that a state change has occurred. 
Finally, it reloads the UI to display the updated states of the entities.

\section{Conclusion}\label{sec:conclusion}
This tool demonstration paper describes the implementation and usage of \tool, a framework for predicting natural scenarios for home automation to enable the testing of security and safety solutions built for the smart home.
Further, we develop and describe \tools extension to the \homeassistant platform, \toolHA, which allows end-users to generate diverse scenarios with various model parameters (\eg varying the $n$ or the prediction flavor), and automatically execute the scenarios as test cases with real and virtual devices.

\begin{acks}
The authors have been supported in part by the NSF-CNS-2132281 grant. The authors acknowledge the contributions from Amit Seal Ami for his help in finalizing the artifact. Moreover, the following undergraduate students from William \& Mary contributed in developing \toolHA{}: Hannah Cummings, Kaitlin Haynal, Kevin Jiao, Jessica Pesso, Kyoko Minamino, Sarah Wang, and Connor Yu. 
\end{acks}

\bibliographystyle{ACM-Reference-Format}
\bibliography{os,semeru,iot,misc,phone,monkeylab,demo}

\end{document}